\documentclass{iopart}
\usepackage{graphicx}
\usepackage{float}
\usepackage{amssymb}

\newcommand{\be}{\begin{equation}} 
\newcommand{\ee}{\end{equation}}
\newcommand{\bc}{\begin{center}}
\newcommand{\ec}{\end{center}}

\begin{document}

\title{Electromagnetic selection rules in the triangular $\alpha-$cluster model of $^{12}$C }
\author{G.Stellin$^1$, L.Fortunato$^{1,2}\footnote{Corresponding author: {\tt fortunat@pd.infn.it}}$ and A.Vitturi$^{1,2}$}
\address{
1) Dipartimento di Fisica e Astronomia ``G.Galilei'' - Universit\`a di Padova \\
2) I.N.F.N. - Sez. di Padova,\\ 
via F.Marzolo 8,  I-35131 Padova, Italy }

\begin{abstract}
After recapitulating the procedure to find the bands and the states occurring in the $\mathcal{D}_{3h}$ alpha-cluster model of $^{12}$C in which the clusters are placed at the vertexes of an equilateral triangle, we obtain the selection rules for electromagnetic transitions. 
While the alpha cluster structure leads to the cancellation of E1
transitions, the approximations carried out in deriving the roto-vibrational hamiltonian
lead to the disappearance of M1 transitions. Furthermore, although in general the lowest active modes are E2, E3, $\cdots$ and M2, M3, $\cdots$,
 the cancellation of M2, M3 and M5 transitions between certain bands also occurs, as a result of the application
of group theoretical techniques drawn from molecular physics. These implications can be very relevant for the spectroscopic analysis of $\gamma$-ray spectra of $^{12}$C.
\end{abstract}
\maketitle

\section{Historical background}
The history of alpha clusters in nuclei started very early in nuclear physics, mainly with the works of J.A.Wheeler \cite{Whe37bis, Whe37}. As an alternative to the Hartree-Fock method, he introduced a molecular viewpoint in nuclear structure, according to which the nucleons are distributed in groups of protons and neutrons of variable composition in the course of time, and this is reflected in the eigenfunctions of the system, that are made up of properly antisymmetrized products of some parts of the single-particle microscopic wavefunctions. Of course the partitions that correspond to stable clusters are favoured in the expansion of the total wavefunction into all possible partitions and, among these, clusters with higher binding energy are even more favoured. This view is corroborated by the well-known fact that nuclear forces saturate, i.e. the  binding of nucleons in nuclei is essentially due to the interaction with the closest neighbours. Therefore clusterization into smaller aggregates might be favoured over other configurations, when the clusters are tightly bound. If the diffusion time of a nucleon inside a cluster is long enough to be comparable or larger than the typical vibrational time between clusters, then these might preserve their structure long enough to make possible a description in terms of rotations and vibrations of clusters around equilibrium configurations \cite{Whe37}. In particular when they are $\alpha$ particles and we concentrate on $\alpha-$conjugate nuclei, the clusters posses such a high binding energy that the description in terms of roto-vibrations of alpha particles placed at the vertexes of a highly symmetric structure of equilibrium points becomes possible and even energetically favoured and this is true as long as one confines the analysis to energies lower than the internal excitation energies of the clusters. This was recognized as early as 1937 by Wheeler in \cite{Whe37}. Just a year later, Hafstad and Teller \cite{HTe38} found that the binding energy of $\alpha-$conjugate nuclei shows a linear behaviour with respect to the number of bonds in ordered geometrical molecular structures. 
More recently the alpha cluster model has been very conveniently reshaped into the Algebraic Cluster Model (ACM) by Bijker and Iachello \cite{BIa00}, where, similarly to what happens with the Vibron model, the bilinear products of boson creation and annihilation operators are associated with the relative motion of clusters and define unitary algebras that are used as spectrum generating algebras. In addition, the ACM imposes discrete symmetries on top of the algebraic description, thus ensuring that molecular types of rotovibrations are properly taken into account. This model is described in a series of papers \cite{Bij02, Bij03,Bij14} that follow the initial application to molecules with $X_3$ structure \cite{BDL95}.
The success of the ACM in reproducing part of the observed energy spectrum of $^{12}$C \cite{BIa00} and the predictions that the model offers have pushed the experimental activity in this field. The group of Birmingham, in particular, has re-examined several older data and made new measurements that add some excited states to the list of cluster states \cite{MLB14}. See also the recent proceedings \cite{Gai2015, Gai2015a, Kok14-01, Kok14-02}.
Most of the cited works have concentrated on the energy spectrum, while little attention has been paid to the electromagnetic transitions between alpha-cluster states. This is a crucial piece of the puzzle, because, {\it contrarily to expectations}, the selection rules are different from those that one could expect based on the spherical shell model or on the collective model: the mere fact that here some discrete point-group symmetry has been imposed on the theoretical framework, i.e.:
$$
\cdots \supset SO(3) \supset \mathcal{D}_{3h} \;,
$$
 implies that the selection rules will follow special patterns, that are usually familiar to molecular physicists, but often fall outside the toolbox of nuclear physicists that are used to selection rules based on angular momentum descending from the continuous spherical rotation group, i.e.: $$ \cdots \supset SO(3) \supset SO(2)\;.$$
The main aim of the present work, that is partly based on the master thesis of one of the authors \cite{Stellin15}, is to discuss the electromagnetic operators and the selection rules in the present context. To the best of our knowledge this crucial piece of information is missing from the literature on this topic. 
The general method that we apply in the following is to find the behaviour of the rotations and of the normal modes of vibrations under the transformations of the $\mathcal{D}_{3h}$ group. This is used to find the admissible states for 3 bosons placed at the vertexes of an equilateral triangle (Sect. \ref{sec2}). 
Eventually, we discuss the way in which the e.m. operators transform (Sect. \ref{sec3}) and we find the selection rules in Sect. \ref{sec4}.
In Sect. \ref{sec5} we discuss a few points that can be very relevant for the spectroscopic analysis of $\gamma$-ray spectra of $^{12}$C.

\section{$^{12}$C: three $\alpha$ particles at the vertexes of an equilateral triangle}
\label{sec2}
The complete point-group symmetry of a system of three spinless bosons with equilateral triangular symmetry is $\mathcal{D}_{3h}$, the 12 elements of which can be grouped into six classes of operations, indicated in the Tables \ref{Tab:powers} and \ref{Tab:roto-states} below \cite{Car}.
This group has a number of possible subgroups among which one can mention for example $D_3$ and $S_3$ \cite{Bij95, Car}. 
The irreducible representations of $\mathcal{D}_{3h}$ can be correlated (by a process of descent in symmetry) with that of smaller groups, at the price of losing some information. In fact polar and axial vectors as well as symmetric quadrupole tensors transform according to certain irreps and give precise selection rules. These irreps somewhat coalesce in the process of descent in symmetry and the selection rules becomes less stringent. In other words the largest group is richer and allows a finer discrimination of certain features. While some authors, for the sake of simplicity, prefer to use the representation labels coming from these 
subgroups, we must adhere here to the notation coming from the full group.
 
\begin{figure*}
\includegraphics[clip=,width=1\textwidth, bb=100 540 580 700]{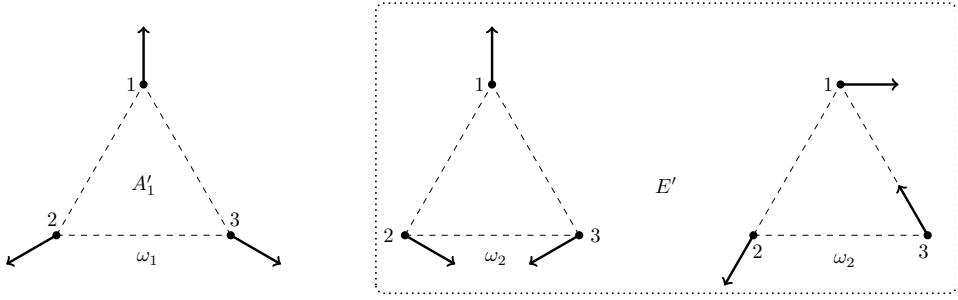}
\caption{Graphical representation of normal modes of vibration. The non-degenerate normal mode ($\omega_1$) that transforms as the completely symmetric representation $A_1'$ consists in an isotropic breathing mode, while the doubly degenerate mode ($\omega_2$) that transforms as $E'$ can be identified as
 a bending mode.}\label{fig:modes}
\end{figure*}

In the reference frame rotating with the nucleus (or intrinsic), let us define as $\Delta x_i, \Delta y_i$ and $\Delta z_i$ the displacements of the i-th $\alpha$ particle with respect to the vertexes (equilibrium points) of the equilateral triangle. With reference to Fig. (\ref{fig:modes}), the intrinsic coordinates for the normal vibrational modes can be expressed as:
$(Q_1, Q_2, Q_3)$, 
\begin{equation}
Q_1 = -\frac{\sqrt{m}}{\sqrt{3}}\Delta y_1+ \frac{\sqrt{m}}{2}\Delta x_2 + \frac{\sqrt{m}}{2\sqrt{3}}\Delta y_2-\frac{\sqrt{m}}{2}\Delta x_3 + \frac{\sqrt{m}}{2\sqrt{3}}\Delta y_3\label{eqn:s18}
\end{equation}
\begin{eqnarray}
Q_2 &=& \left(-\frac{1}{2}+\frac{1}{\sqrt{3}}\right)\frac{\sqrt{m}}{\sqrt{6}}\Delta x_1- \left(\frac{1}{2}+\frac{1}{\sqrt{3}}\right)\frac{\sqrt{m}}{\sqrt{6}}\Delta y_1  \nonumber\\
&-& \left(\frac{1}{2}+\frac{1}{\sqrt{3}}\right)\frac{\sqrt{m}}{\sqrt{6}}\Delta x_2 + \left(-\frac{1}{2}+\frac{1}{\sqrt{3}}\right)\frac{\sqrt{m}}{\sqrt{6}}\Delta y_2 \nonumber \\
&+&\frac{\sqrt{m}}{\sqrt{6}}\Delta x_3 + \frac{\sqrt{m}}{\sqrt{6}}\Delta y_3 \label{eqn:s19}
\end{eqnarray}
\begin{eqnarray}
Q_3 &=& \left(\frac{1}{2}+\frac{1}{\sqrt{3}}\right)\frac{\sqrt{m}}{\sqrt{6}}\Delta x_1 + \left(-\frac{1}{2}+\frac{1}{\sqrt{3}}\right)\frac{\sqrt{m}}{\sqrt{6}}\Delta y_1 \nonumber \\
&+& \left(\frac{1}{2}-\frac{1}{\sqrt{3}}\right)\frac{\sqrt{m}}{\sqrt{6}}\Delta x_2-\left(\frac{1}{2}+\frac{1}{\sqrt{3}}\right)\frac{\sqrt{m}}{\sqrt{6}}\Delta y_2
\nonumber \\
&-&\frac{\sqrt{m}}{\sqrt{6}}\Delta x_3 + \frac{\sqrt{m}}{\sqrt{6}}\Delta y_3, \label{eqn:s20}
\end{eqnarray}

The reducible representation of the cartesian displacements can be found from the character of each of the transformation matrices. From this, by subtracting the characters of the representations under which the rotation and translation coordinates\footnote{With $I^e_{\alpha\alpha}$ we indicate the moment of inertia with respect to the $\alpha \in x,y,z$ axis of the intrinsic reference frame.} \cite{Stellin15},
\be
T_\alpha = \sqrt{3m} \sum_{i=1}^3 m\Delta\alpha_i
\ee
and 
\be
R_\alpha = \sqrt{I^e_{\alpha\alpha}}\sum_{i=1}^3\sum_{\beta,\gamma=x,y,z} m \epsilon_{\alpha\beta\gamma} \beta_i^e\Delta\gamma_i \;,
\ee
transform, one arrives at the characters for the normal coordinates that can be decomposed into $\Gamma(Q_1,Q_2,Q_3)= A_1'\oplus E'$ as represented in Fig.(\ref{fig:modes}), that is a direct sum of a one-dimensional representation, $A_1'$, with coordinate $Q_1$ and a two-dimensional representation, $E'$, with coordinates $(Q_2,Q_3)$.

Therefore, assuming decoupled harmonic motion for each mode and a rigid oblate symmetric top for rotations, the generic rotovibrational states of $^{12}$C$\equiv (3\alpha)$ can be written as:
\be
\mid n_1, n_2, J,K,M \rangle = \Phi_{n_1}^A(Q_1) \Phi_{n_2}^E(Q_2,Q_3) \psi_{JKM}
\label{fact}
\ee
where $n_1$ and $n_2$ are the phonon numbers in the respective vibrations, and $\psi_{JKM}$ is the rotational wavefunction. These states are eigenstates of the hamiltonian with harmonic vibrations plus rotations around equilibrium points, but they can also be used as a possible O.N. basis for more complex hamiltonian operators with generic potentials.  
In group theoretical notation, we can rewrite the factorized wavefunctions in terms of their representations under transformations of the group
\be
 \underbrace{[\Gamma(A_1')]^{n_1} \otimes [\Gamma(E')]^{n_2}}_{vib.} \otimes \underbrace{[\Gamma( D^{(J)}_{KM} )]}_{rot.}
\ee

Any power of the totally symmetric mode will preserve the same character,because any multiplication of $A_1'$ with itself will give $A_1'$ again, but the character of multiphonon vibrations of the doubly degenerate type must be determined according to the recursive formula for symmetric powers of an irreducible 2-dimensional representation $\Gamma[R]$ (proven in Ref.\cite{Ham}) 
\be
\chi^{\Gamma^{(v)}} \big[ R \big] = \frac{1}{2} \Bigl( \chi^{\Gamma}\big[R\big]\chi^{\Gamma^{(v-1)}}\big[R\big] +\chi^{\Gamma^{(v)}}\big[R^2\big] \Bigr)
\ee
where $R \in \mathcal{D}_{3h}$ is an operation of the group and $\Gamma$ is either $E'$ or $E''$. 
The final result, given also in Ref. \cite{Stellin15} is summarized in Table \ref{Tab:powers}.

\begin{table}[ht]
\begin{center}
\begin{tabular}{ccccccc}
\hline\hline
$\mathcal{D}_{3h}$ & $\mathbb{I}$ & $2C_3$ & $3C_2$ & $\sigma_h$ & $2S_3$ & $3\sigma_V$\\
\hline
$\Gamma(\Phi_1^E)=E'$ & $2$ &-$1$ & $0$ & $2$ & -$1$ & $0$\\
$\Gamma(\Phi_2^E)= A_1'\oplus E'$ & $3$ & $0$ & $1$ & $3$ & $0$ & $1$\\
$\Gamma(\Phi_3^E)= A_1'\oplus A_2'\oplus E'$ &  $4$ & $1$ & $0$ & $4$ & $1$ & $0$\\
$\Gamma(\Phi_4^E)= A_1'\oplus E' \oplus E'$ & $5$ & -$1$ & $1$ & $5$ & -$1$ & $1$\\ 
\hline \hline
\end{tabular}
\end{center}
\caption{Characters of representations for vibrational eigenfunctions of type $\omega_2$ (with $n_2 \leq 4$, see \cite{Stellin15} for  $n_2 \leq 10$).}
\label{Tab:powers}
\end{table}

In particular, starting from the coordinates $Q_i$ one can define their conjugate momenta and promote them to differential operators. With these it is possible to construct creation and annihilation operators that can be associated with a harmonic oscillator hamiltonian. Using the transformation properties of these operators one can prove that, whenever $m$ is not multiple of 3, the pair of eigenfunctions $(\Phi_{n_2,-m}^E, \Phi_{n_2,m}^E)$ transforms under the irrep $E'$, while in the other case the linear combinations: 
\begin{equation}
\frac{\Phi_{n_2, m}^E+(-1)^m \Phi_{n_2,-m}^E}{\sqrt{2}}\hspace{0.5cm}\mathrm{and}\hspace{0.5cm}\frac{\Phi_{n_2, m}^E-(-1)^m \Phi_{n_2,-m}^E}{\sqrt{2}} \label{eqn:158bis}
\end{equation}
transform respectively under the irreps $A'_1$ and $A_2'$. Here $m$ represents the quantum number that takes the values $m=-n_2, -n_2+2, \cdots, n_2-2, n_2$ and it can be used to distinguish the $(n_2+1)$ degenerate states in the multiplet corresponding to $n_2$ phonons.

\subsection{Character of the rotational part of the wavefunction}
The characters of the rotational part of the wavefunction can also be found by inspecting the character of the wavefunctions under the action of the elements of the rotation group. It is possible to associate an element of the permutation-inversion group $\mathcal{S}_3 \times \mathcal{C}_i$, isomorphic to $\mathcal{D}_{3h}$, to a proper rotation of the intrinsic axes, i.e. an element of the three-dimensional rotation group, SO(3). Moreover, in this particular case, it's possible to express the action of an operation of the permutation-inversion group as a rotation about the $z$-axis of the intrinsic frame of angle $\beta$, $R_{z}^{\beta}$, or a rotation of $\pi$ about an axis lying in the plane $xy$ forming an angle $\alpha$ with the $x$-axis of the same frame of reference, $R_{\alpha}^{\pi}$. Given that the rotational part of the total wavefunction can be chosen as a normalized Wigner D matrix, 
\begin{equation}
\psi_{JKM} \equiv \sqrt{\frac{2J+1}{4\pi^2}} D^{J*}_{MK}(\varphi, \theta, \chi).\nonumber
\end{equation}
the symmetry character of each eigenfunction characterized by $J$, $K$ and $M$ (see Tab. \ref{Tab:roto-states}) under the operations of the permutation-inversion group can be obtained by making use of the following relations
\begin{equation}
R^{\beta}_z\psi_{JKM} = e^{iK\beta}\psi_{JKM} \nonumber
\end{equation}
\begin{equation}
R^{\pi}_{\alpha}\psi_{JKM} =(-1)^J e^{-2iK\alpha}\psi_{J-KM}. \nonumber
\end{equation}

\begin{table}[ht]
\begin{center}
\begin{small}
\begin{tabular}{c|c|c|cccccc}
\hline \hline
$J$ & $|K|$ & $\mathcal{D}_{3h}$ & $\mathbb{I}$ & $2C_3$ & $3C_2$ & $\sigma_h$ & $2S_3$ & $3\sigma_V$\\
\hline
$0$ & $0$ & $A_1'$ & $1$ & $1$ & $1$ & $1$ & $1$ & $1$\\
\hline
$1$ & $0$ &  $A_2'$ & $1$ & $1$ & -$1$ & $1$ & $1$ & -$1$\\
& $1$ & $E''$ & $2$ & -$1$ & $0$ & -$2$ & $1$ & $0$\\
\hline
$2$ & $0$ & $A_1'$ & $1$ & $1$ & $1$ & $1$ & $1$ & $1$\\
& $1$ & $E''$ & $2$ & -$1$ & $0$ & -$2$ & $1$ & $0$\\
& $2$ & $E'$ & $2$ & -$1$ & $0$ & $2$ & -$1$ & $0$\\
\hline
$3$ & $0$ & $A_2'$ & $1$ & $1$ & -$1$ & $1$ & $1$ & -$1$\\
& $1$ & $E''$ & $2$ & -$1$ & $0$ & -$2$ & $1$ & $0$\\
& $2$ & $E'$ & $2$ & -$1$ & $0$ & $2$ & -$1$ & $0$\\
& $3$ & $A_1''\oplus A_2''$ & $2$ & $2$ & $0$ & -$2$ & -$2$ & $0$\\
\hline
$4$ & $0$ & $A_1'$ & $1$ & $1$ & $1$ & $1$ & $1$ & $1$\\
& $1$ & $E''$ & $2$ & -$1$ & $0$ & -$2$ & $1$ & $0$\\
& $2$ & $E'$ & $2$ & -$1$ & $0$ & $2$ & -$1$ & $0$\\
& $3$ & $A_1''\oplus A_2''$ & $2$ & $2$ & $0$ & -$2$ & -$2$ & $0$\\
& $4$ & $E'$ & $2$ & -$1$ & $0$ & $2$ & -$1$ & $0$\\
\hline
$5$ & $0$ & $A_2'$ & $1$ & $1$ & -$1$ & $1$ & $1$ & -$1$\\
& $1$ & $E''$ & $2$ & -$1$ & $0$ & -$2$ & $1$ & $0$\\
& $2$ & $E'$ & $2$ & -$1$ & $0$ & $2$ & -$1$ & $0$\\
& $3$ & $A_1''\oplus A_2''$ & $2$ & $2$ & $0$ & -$2$ & -$2$ & $0$\\
& $4$ & $E'$ & $2$ & -$1$ & $0$ & $2$ & -$1$ & $0$\\
& $5$ & $E''$ & $2$ & -$1$ & $0$ & -$2$ & $1$ & $0$\\
\hline
$6$ & $0$ & $A_1'$ & $1$ & $1$ & $1$ & $1$ & $1$ & $1$\\
& $1$ & $E''$ & $2$ & -$1$ & $0$ & -$2$ & $1$ & $0$\\
& $2$ & $E'$ & $2$ & -$1$ & $0$ & $2$ & -$1$ & $0$\\
& $3$ & $A_1''\oplus A_2''$ & $2$ & $2$ & $0$ & -$2$ & -$2$ & $0$\\
& $4$ & $E'$ & $2$ & -$1$ & $0$ & $2$ & -$1$ & $0$\\
& $5$ & $E''$ & $2$ & -$1$ & $0$ & -$2$ & $1$ & $0$\\
& $6$ & $A_1'\oplus A_2'$ & $2$ & $2$ & $0$ & $2$ & $2$ & $0$\\
\hline\hline
\end{tabular}
\end{small}
\end{center}
\caption{Characters of the representations of rotational states up to $J=6$. We note that: (i) for even values (resp. odd) of $J$ the irrep with $K=0$ is $A_1'$ (resp. $A_2'$); (ii) for states with any $J$ and $K$ odd (resp. even) and not multiple of $3$ the irrep is $E''$ (resp. $E'$); (iii) for states with any $J$ and $K$ odd (resp. even) but multiple of $3$ the irrep is $A_1''\oplus A_2''$ (resp. $A_1'\oplus A_2'$), that is a {\it faux} pair.}
\label{Tab:roto-states}
\end{table}

\subsection{Global character of wavefunction}
We are now equipped with the behaviour under transformations of the group elements for all the parts of the factorized wavefunction (\ref{fact}), namely the trivial part corresponding to the breathing mode (that gives always $A_1'$) and the non-trivial parts corresponding to the doubly-degenerate bending mode (Table \ref{Tab:powers}) and rotations (Table \ref{Tab:roto-states}). We can thus proceed to the selection of states, i.e. we can put all the requirements together and reproduce and extend the table given by Wheeler. This is far from trivial and the results are displayed in Table \ref{Tab:states} \footnote{See master thesis of G.Stellin for a bigger table \cite{Stellin15}}, where the numbers count the occurrences of either $A_1'$ or $A_1''$ in the total product of characters. This has to be so because we are looking for bosonic representations, i.e. representations that give +1 under permutation of any two alpha's. The permutations coincide with the $I$, $2C_3$ and $3C_2$ operations of the group and the only irreps that have character +1 for all these are $A_1'$ and $A_1''$ (they differ only under inversion operations, that correspond to parity). 
\begin{table}[ht]
\begin{center}
\begin{tabular}{c|c|c| ccc cc}
\hline \hline 
\multicolumn{3}{c}{$n_1=0,1,2,3,...$; $n_2=$} & $0$ & $1$ & $2$ & $3$ & $4$\\
\hline 
$J$ & $|K|  $ & $\mathcal{D}_{3h}(N)$  & $A_1'$ & $E'$ & $A_1'\oplus E'$ & $A_1'\oplus A_2' \oplus E'$ & $A_1' \oplus E' \oplus E'$ \\ 
\hline 
$0$ & $0$ & $A_1'$ & $1$ & $0$ & $1$ & $1$ & $1$ \\
\hline 
$1$ & $0$ & $A_2'$ & $0$ & $0$ & $0$ & $1$ & $0$ \\
 & $1$ & $E''$ & $0$ & $1$ & $1$ & $1$ & $2$ \\
\hline 
$2$ & $0$ & $A_1'$ & $1$ & $0$ & $1$ & $1$ & $1$ \\
 & $1$ & $E''$ & $0$ & $1$ & $1$ & $1$ & $2$ \\
& $2$ & $E'$ & $0$ & $1$ & $1$ & $1$ & $2$ \\
\hline 
$3$ & $0$ & $A_2'$ & $0$ & $0$ & $0$ & $1$ & $0$ \\
& $1$ & $E''$ & $0$ & $1$ & $1$ & $1$ & $2$ \\
& $2$ & $E'$ & $0$ & $1$ & $1$ & $1$ & $2$ \\
& $3$ & $A_1''\oplus A_2''$ & $1$ & $0$ & $1$ & $2$ & $1$\\
\hline 
$4$ & $0$ & $A_1'$ & $1$ & $0$ & $1$ & $1$ & $1$ \\
 & $1$ & $E''$ & $0$ & $1$ & $1$ & $1$ & $2$ \\
& $2$ & $E'$ & $0$ & $1$ & $1$ & $1$ & $2$ \\
& $3$ & $A_1''\oplus A_2''$ & $1$ & $0$ & $1$ & $2$ & $1$\\
& $4$ & $E'$ & $0$ & $1$ & $1$ & $1$ & $2$\\
\hline 
$5$ & $0$ & $A_2'$ & $0$ & $0$ & $0$ & $1$ & $0$ \\
& $1$ & $E''$ & $0$ & $1$ & $1$ & $1$ & $2$ \\
& $2$ & $E'$ & $0$ & $1$ & $1$ & $1$ & $2$ \\
& $3$ & $A_1''\oplus A_2''$ & $1$ & $0$ & $1$ & $2$ & $1$ \\
& $4$ & $E'$ & $0$ & $1$ & $1$ & $1$ & $2$ \\
& $5$ & $E''$ & $0$ & $1$ & $1$ & $1$ & $2$ \\
\hline 
$6$ & $0$ & $A_1'$ & $1$ & $0$ & $1$ & $1$ & $1$ \\
& $1$ & $E''$ & $0$ & $1$ & $1$ & $1$ & $2$ \\
& $2$ & $E'$ & $0$ & $1$ & $1$ & $1$ & $2$ \\
& $3$ & $A_1''\oplus A_2''$ & $1$ & $0$ & $1$ & $2$ & $1$ \\
& $4$ & $E'$ & $0$ & $1$ & $1$ & $1$ & $2$ \\
& $5$ & $E''$ & $0$ & $1$ & $1$ & $1$ & $2$ \\
& $6$ & $A_1'\oplus A_2'$ & $1$ & $0$ & $1$ & $2$ & $1$ \\
\hline \hline 
\end{tabular}
\end{center}
\caption{Allowed roto-vobrational states for three bosons at the vertexes of an equilateral triangle, $\forall n_1$, with $\Phi_{n_2}^E$ up to $n_2=4$ and $\psi_{JKM}$ up to $J=6$.}
\label{Tab:states}
\end{table}
This table displays the sequence of states that are expected in each band, for example in the ground state band we should expect all the states contained in the first column, i.e. $0^+, 2^+, 3^-, 4^\pm, 5^-, 6^{\pm,+}, \cdots$ and explains also the somewhat puzzling parities. In fact all the states with even $K$ have positive parity, while states with odd $K$ have negative parity. For larger values of $J$, the number of allowed states increases and some of the parities are repeated. This fact can be seen in the following way: take the first column, the product of $A_1'$ for the vibrational state with $n_2=0$ with the rotational states gives exactly the characters of the rotational states, therefore we sign one every time we see an $A_1'$ or an $A_1''$ in the column. Hence, despite the band having a global $A_1'$ character the full rotovibrational states have either $A_1'$ or an $A_1''$ corresponding to their parities. Indeed $A_1'$ appears only for even values of $K$ and $A_1''$ appears only for odd values. Take now the second column with the sequence of states $1^-, 2^\pm, 3^\pm,4^{\pm,+}, 5^{\pm,\pm}, 6^{\pm, \pm}, \cdots $: this band has a vibrational $E'$ character, but that does not mean that all its states have it. In fact the multiplication with the rotational characters determines where and how many $A_1'$ and $A_1''$ states appears. For instance the $J=5$, $K=1$ state is obtained by multiplying $E''$ with $E'$  and this gives $A_1''\oplus A_2'' \oplus E''$ according to the multiplication tables of the $\mathcal{D}_{3h}$ group. This means that there is only one state with negative parity. Further inspection on the transformation properties of the whole rotovibrational states leads, finally, to the form of the allowed eigenfunctions, enumerated in Tab. \ref{Tab:states}: for the states with $n_2 = 0$ with $|K|$ not multiple of 3, 
\begin{equation}
\Phi_{n_1}^A\Phi_{0,0}^E\left(\frac{\psi_{JKM}+(-1)^{K+J}\psi_{J-KM}}{\sqrt{2}}\right),\nonumber
\end{equation}
else 
\begin{equation}
\Phi_{n_1}^A\Phi_{0,0}^E\psi_{J0M}.\nonumber
\end{equation}
On the other hand, when $n_2\neq 0$, the admitted eigenfunctions preserve the rotational-vibrational factored form when $\mid m\mid$ and $\mid K\mid$ are multiple of 3, 
\begin{equation}
\Phi_{n_1}^A\left(\frac{\Phi_{n_2,m}^E+(-1)^{m}\Phi_{n_2,-m}^E}{\sqrt{2}}\right)\left(\frac{\psi_{JKM}+(-1)^{K+J}\psi_{J-KM}}{\sqrt{2}}\right),\nonumber
\end{equation}
but they lose this nature when the two quantum numbers are not divisible by 3,
 \begin{equation}
\Phi_{n_1}^A\left(\frac{\Phi_{n_2,m}^E\psi_{JKM}+(-1)^{J+K+m}\Phi_{n_2,-m}^E\psi_{J-KM}}{\sqrt{2}}\right)\nonumber.
\end{equation}
The final form of these states is, in fact, useful for the explicit computation of the reduced transition probabilities of electric or magnetic multipole B(R$\lambda$, $|i\rangle \rightarrow |f\rangle$) with $R=E,M$ or the intrinsic electric quadrupole moments of the allowed states.

\section{E.M. operators}\label{sec3}
The electric (E) and magnetic (M) multipole operators, 
$\Omega_{\lambda\mu}(R)$, 
that make the transitions between two nuclear rotovibrational states possible, 
are irreducible spherical tensors of rank $\lambda$, i.e. objects that transform under a rotation of angles $(\varphi, \theta, \chi)$ in the following way,
\begin{equation}
\Omega_{\lambda\mu} (R) = \sum_{\nu=-\lambda}^{\lambda} D_{\mu\nu}^{\lambda *}(\varphi, \theta, \chi)\omega_{\lambda\nu}(R)
\end{equation}
where $R=E,M$ and where the intrinsic electric multipole operators, $\omega_{\lambda\nu}(E)$ can be expressed solely in terms of displacements from the equilibrium positions of the $\alpha$ particles in the intrinsic frame and $D_{\mu\nu}^\lambda$ are the Wigner matrices.
The operators $\Omega_{\lambda\mu}(R)$ and $\omega_{\lambda\mu}(R)$ possess different transformation properties under the operations of the permutation-inversion group $\mathcal{D}_{3h}$: we are going to inspect their symmetry characters separately for the electric and magnetic multipole operators.
The charge density operator $\hat{\rho}(\vec{r})$ of the alpha clusters is defined as
\begin{equation}
\hat{\rho}(\vec{r})=2e\sum_{i=1}^3 \delta(\vec{r}-\vec{r}_i),
\end{equation} 
and the expression of the intrinsic electric multipole operators becomes
\begin{equation}
\omega_{\lambda\nu}(E) = \int \mathrm{d}^3r \hspace{1mm} r^{\lambda} Y_{\lambda\nu}(\varphi, \theta)\hspace{1mm}\hat{\rho}(\vec{r}) \;.
\end{equation} 

Therefore we can analyze the symmetry characters of these operators in the $\mathcal{D}_{3h}$ group, obtaining the decomposition presented in Table \ref{Tab:oper}.  This process, that soon becomes cumbersome, can be simplified following the guidelines presented in Ref. \cite{Gtw95}, namely finding at first that the electric dipole transforms as a polar vector, i.e. as $A_2'' \oplus E'$ and then observing that the $\lambda$-th electric multiple reducible representation is the symmetric $\lambda$-th power of the three dimensional reducible representation $\Gamma$ of $\omega_{1\mu}(E)$, 
\begin{equation}
\chi^{\Gamma^{(\lambda)}}[R]=\end{equation}
$$
=\frac{1}{3}\left[2\chi^{\Gamma}[R]\chi^{\Gamma^{(v-1)}}[R] +\frac{1}{2}\left(\chi^{\Gamma}[R^2]-\chi^{\Gamma}[R]^2\right)\chi^{\Gamma^{(v-2)}}[R]+\chi^{\Gamma}[R^v]\right] \;, 
$$
with $R \in \mathcal{D}_{3h}$, and then subtracting the representation of the same parity and multipolarity of lower order. Furthermore, the choice of the $z$ axis of the intrinsic triad allows to associate, with further inspection, each irreducible representation (or couple of irreps) appearing in the decomposition of the representation of the full intrinsic electric $2^{\lambda}$-pole operators to a particular pair, $\omega_{\lambda,\pm |\nu|}(E)$, that becomes a singlet when $\nu=0$: the result of this analysis is reported in Table \ref{Tab:electrocompo}. 
Notice that the brute-force algorithm to fill up this table can be stated as follows: starting from the known character of dipole, construct the quadrupole as a double application of dipole, component by component, the product of representations with $(\lambda=1, \mu_1)$ and $(\lambda=1, \mu_2)$ will give the character of $(\lambda=2, \mu_1+\mu_2)$, taking care that in the product of a representation with itself only the symmetric part must be retained. All $\mu=0$ must be singlets, while all others must be two-dimensional. Then proceed recursively to higher multipolarities, keeping in mind the observation for $\mu=3k$.   
  
\begin{table}[ht]
\begin{center}
\begin{tabular}{c|ccccccc}
\hline
\small{$|\mu|$} $\rightarrow$ & $0$ & $1$ & $2$ & $3$ & $4$ & $5$ & $6$ \\
\small{$\lambda$}$\downarrow$ & & & & & & & \\
\hline
\small{$0$} & \small{$A_{1}'$} &  & &  &  & & \\
\small{$1$} & \small{$A_2''$} & \small{$E'$} &  &  &  & & \\
\small{$2$} & \small{$A_{1}'$} & \small{$E''$} & \small{$E'$} & & & & \\
\small{$3$} & \small{$A_2''$} & \small{$E'$} & \small{$E''$} & \small{$A_1'+A_2'$} & & & \\
\small{$4$} & \small{$A_1'$} & \small{$E''$} & \small{$E'$} & \small{$A_1''+A_2''$} & \small{$E'$} & & \\
\small{$5$} & \small{$A_2''$} & \small{$E'$} & \small{$E''$} & \small{$A_1'+A_2'$} & \small{$E''$} & \small{$E'$} & \\
\small{$6$} & \small{$A_1'$} & \small{$E''$} & \small{$E'$} & \small{$A_1''+A_2''$} & \small{$E'$} & \small{$E''$} & \small{$A_1'+A_2'$}\\
\hline
\end{tabular}
\caption{Transformation properties of the intrinsic electric multipole operators $\omega_{\lambda\mu}(E)$ with $\lambda \leq 6$. As we can infer, for even multipolarity $\lambda$ if $|\mu|\neq 0$ is even (resp. odd) and not multiple of 3 it transforms according to the irreps $E'$ ($E''$), while if it's even (resp. odd) and multiple of three it behaves as $A_1'+A_2'$ ($A_1''+A_2''$) and if it's zero then it transforms like $A_1'$. For odd multipolarity, when $|\mu|\neq 0$ is even (odd) and not multiple of 3 it transforms according to the irreps $E''$ ($E'$), while if it's even (odd) and multiple of 3 it behaves as $A_1''+A_2''$ ($A_1'+A_2'$) and if it's zero then it transforms like $A_2''$.}\label{Tab:electrocompo}
\end{center}

\end{table}

\begin{table}[ht]
\begin{center}
\begin{tabular}{c|cccccc}
\hline \hline
$\mathcal{D}_{3h} (N)$ & $\mathbb{I}$ & $2C_3$ & $3C_2$ & $\sigma_h$ & $2S_3$ & $3\sigma_V$\\
\hline
$\Gamma[\omega_{00}(E)]= A_1'$ & $1$ & $1$ & $1$ & $1$ & $1$ & $1$\\
$\Gamma[\omega_{1\mu}(E)] = A_2''+ E'$ & $3$ & $0$ & -$1$ & $1$ & -$2$ & $1$\\
$\Gamma[\omega_{2\mu}(E)]= A_1' + E' + E''$ & $5$ & -$1$ & $1$ & $1$ & $1$ & $1$\\
$\Gamma[\omega_{3\mu}(E)] = A_1' + A_2'+ A_2'' + E' + E''$ & $7$ & $1$ & -$1$ & $1$ & $1$ & $1$\\
$\Gamma[\omega_{4\mu}(E)] = A_1'+ A_1'' + A_2'' +2 E' + E''$ & $9$ & $0$ & $1$ & $1$ & -$2$ & $1$\\
$\Gamma[\omega_{5\mu}(E)] = A_1'+ A_2' + A_2'' +2 E' +2 E''$ & $11$ & -$1$ & -$1$ & $1$ & $1$ & $1$\\
$\Gamma[\omega_{6\mu}(E)] = 2A_1' + A_2' + A_1'' + A_2'' +2 E' +2 E''$ & $13$ & $1$ & $1$ & $1$ & $1$ & $1$\\
\hline \hline
\end{tabular}
\end{center}
\caption{Character table for the electric multipole operators $\omega_{\lambda\mu}(E)$ with $\mu=0,\pm1,...,\pm\lambda$ e $\lambda \leq 6$ \cite{Gtw95} under transformations of the $\mathcal{D}_{3h}$ group. These properties are fundamentally different from those obtained in the laboratory frame. Moreover, we note that the dipole operator has the same transformation properties (same irreps) of the $T_{\alpha}$ coordinates, a fact that, as we shall see in the following, is not without consequences.}
\label{Tab:oper}
\end{table}
Notice that the content of Table \ref{Tab:oper} coincide with the  $\mathcal{D}_{3h}$ line of Table 1 and 2 of Ref. \cite{Gtw95}, except for the octupole\footnote{The published version \cite{Gtw95} contains a few incorrect entries, namely the 32- and 64-poles for $\mathcal{D}_{5}$ and the octupole for $\mathcal{D}_{3h}$. We confirmed this fact with the authors that knew it since soon after the publication (A. Gelessus, private communication, 5/11/2015). The on-line version of the same table, that can be found at http://symmetry.jacobs-university.de, contains the correct decompositions.}.

A similar analysis can be carried out for the magnetic case. Starting from the density current
\be
\hat{\vec{j}}(\vec{r}) = \frac{2e}{m_\alpha} \sum_{j=1}^3 \vec p_i \delta(\vec{r}-\vec{r}_i)
\ee
we can see that the behaviour of the magnetic multipole operator is that of a tensor of rank $\lambda$ (through relation 17.19 in Ref. \cite{deSh}):
\begin{equation}
\omega_{\lambda\nu}(M) =  \frac{2\hbar c}{m_\alpha c^2(\lambda+1)} \int \mathrm{d}^3r \hspace{1mm}\hat{\vec{L}} \cdot \vec \nabla \bigl(r^\lambda Y_{\lambda\nu}(\varphi, \theta)\bigr)
\end{equation}
where the magnetization term is zero because $\alpha$ particles are spinless bosons. 

Differently to the preceeding case, the magnetic dipole operator behaves as an axial vector, transforming as the $A_2'\oplus E''$ reducible representation under the operations of $\mathcal{D}_{3h}$ like the   components of the angular momentum $R_{i}$ with $i=x,y,z$. 

\begin{table}[ht]
 \begin{center}
\begin{tabular}{c|ccccccc}
\hline
\small{$|\mu|$} $\rightarrow$ & $0$ & $1$ & $2$ & $3$ & $4$ & $5$ & $6$ \\
\small{$\lambda$}$\downarrow$ & & & & & & & \\
\hline
\small{$0$} & \small{$A_{1}''$} &  & &  &  & & \\
\small{$1$} & \small{$A_2'$} & \small{$E''$} &  &  &  & & \\
\small{$2$} & \small{$A_{1}''$} & \small{$E'$} & \small{$E''$} & & & & \\
\small{$3$} & \small{$A_2'$} & \small{$E''$} & \small{$E'$} & \small{$A_1''+A_2''$} & & & \\
\small{$4$} & \small{$A_1''$} & \small{$E'$} & \small{$E''$} & \small{$A_1'+A_2'$} & \small{$E''$} & & \\
\small{$5$} & \small{$A_2'$} & \small{$E''$} & \small{$E'$} & \small{$A_1''+A_2''$} & \small{$E'$} & \small{$E''$} & \\
\small{$6$} & \small{$A_1''$} & \small{$E'$} & \small{$E''$} & \small{$A_1'+A_2'$} & \small{$E''$} & \small{$E'$} & \small{$A_1''+A_2''$} \\
\hline
\end{tabular}
\caption{Transformation properties of the intrinsic magnetic multipole operators $\omega_{\lambda\mu}(M)$ with $\lambda \leq 6$. 
Rules similar to the preceding case apply. Notice that, for the sake of mathematical completeness, the magnetic monopole has been included as a null pseudoscalar, although clearly there is no physical significance in the monopole operator.}\label{Tab:magneto-compo}
\end{center}

\end{table}

Although the procedure to determine the behaviour of the $\omega_{\lambda\mu}(M)$ is different from that of the intrinsic electric multipole operators, the transformation properties of the former can be recovered consistently with the parity relationship $\pi[\Omega_{\lambda\mu}(M)] = ( -1)^{1}\pi[\Omega_{\lambda\mu}(E)]$ \footnote{This is equivalent to saying that magnetic multipole operators of odd (resp. even) $\lambda$ transform as $A_1'$ (resp. $A_1''$) and viceversa for the electric multipole operators.} by inverting the parity (i.e. exchanging ' with '') of the irreps of the permutation-inversion group, $\mathcal{D}_{3h}$, occurring in the decomposition of the reducible representations of the $\omega_{\lambda\mu}(E)$ in Table \ref{Tab:Moper}.
\begin{table}[ht]
\begin{center}
\begin{tabular}{c|cccccc}
\hline \hline
$\mathcal{D}_{3h} (N)$ & $\mathbb{I}$ & $2C_3$ & $3C_2$ & $\sigma_h$ & $2S_3$ & $3\sigma_V$\\
\hline
$\Gamma[\omega_{1\mu}(M)] = A_2'+ E''$ & $3$ & $0$ & -$1$ & -$1$ & $2$ & -$1$\\
$\Gamma[\omega_{2\mu}(M)]= A_1'' + E' + E''$ & $5$ & -$1$ & $1$ & -$1$ & -$1$ & -$1$\\
$\Gamma[\omega_{3\mu}(M)] =  A_2'+A_1''+ A_2'' + E' + E''$ & $7$ & $1$ & -$1$ & -$1$ & -$1$ & -$1$\\
$\Gamma[\omega_{4\mu}(M)] = A_1' + A_2'+ A_1'' + 2E' + E''$ & $9$ & $0$ & $1$ & -$1$ & $2$ & -$1$\\
$\Gamma[\omega_{5\mu}(M)] =  A_1''+ A_2'+ A_2'' + 2E' + 2E''$ & $11$ & -$1$ & -$1$ & -$1$ & -$1$ & -$1$\\
$\Gamma[\omega_{6\mu}(M)] = A_1'+2A_1''+A_2' + A_2'' + 2E' + 2E''$ & $13$ & $1$ & $1$ & -$1$ & -$1$ & -$1$\\
\hline \hline
\end{tabular}
\end{center}
\caption{Character table for the magnetic multipole operators $\omega_{\lambda\mu}(M)$ with $\mu=0,\pm1,...,\pm\lambda$ e $\lambda \leq 6$ under transformations of the $\mathcal{D}_{3h}$ group. The dipole operator has the same transformation properties (same irreps) of rotations, and therefore it's not an active mode.}
\label{Tab:Moper}
\end{table}
This last table shows that magnetic quadrupole and octupole modes (and all the higher multipoles) have at least one component transforming as $E'$ and thus they are allowed (they can connect states) between A-type and E-type bands or between 2 E-type bands. Slightly different is the case of transitions between A-type bands. Strictly speaking, since $A_1'$ is present only in M4 and M6, two distinct A-type bands with different phonon numbers can only be connected by these multipolarities.

\section{Allowed and forbidden E.M. transitions}
\label{sec4}
Having presented and selected the rotovibrational states with the correct exchange symmetry within
the framework of Wheeler's model, it is interesting to obtain the selection rules that govern the reduced electric and magnetic multipole transition probabilities, $B(R\lambda; \mid i \rangle \rightarrow \mid f \rangle)$, between any two $\alpha-$cluster states of $^{12}$C. These depend on the matrix elements
\be
M_{ij}=\langle \Psi_i^{(J_i)} \mid \Omega_{\lambda \mu} (R)\mid \Psi_j^{(J_j)}\rangle 
\ee
where $R=E,M$.
As previously anticipated, part of these rules owe to the $\mathcal{D}_{3h}$ point group and will be discussed later, while others are related to the continuous symmetry groups, $U(3)\supset SO(3)$, to which the operators in the rotovibrational hamiltonian belong. Accordingly, the above wavefunctions $\Psi_i^{(J_i)}$ can be rewritten also as $\mid J_i K_i M_i \rangle \mid n_{1i} n_{2i} m_{i} \rangle$
and these selection rules can be fully expressed in terms of the
quantum numbers, $(J,K)$, $(n_2,m)$ and $n_1$, which are linked to the preceding Lie groups as follows:
\begin{table}[h!]
\begin{center}
\begin{tabular}{ccccccc}
U(1) & $\otimes $  & U(2) & $\otimes $ & SO(3) & $\supset $ SO(2) & $\supset \mathcal{D}_{3h}$ \\
$\downarrow$ & & $\downarrow$ & & $\downarrow$ & & \\
$n_1$ && $(n_2,m)$ && $(J,K)$ &&\\
\end{tabular}
\end{center}
\end{table}

Let's start the analysis of these conditions from those arising from the vibrational part of the matrix element. As it can be observed by expressing the $\omega_{\lambda\mu}(R)$ in terms of the normal coordinate operators $Q_1,Q_2,Q_3$ and then to the associated second quantization operators, the non-null matrix elements of electric or magnetic 2$^\lambda$-pole must fulfill:
\be
\Delta n = (n_{1f}+n_{2f}) - (n_{1i}+n_{2i}) \in \{ 0,\pm 1, \pm 2,\cdots, \pm \lambda\}
\label{n-rule}
\ee
Since also the degenerate eigenfunctions, $\Phi_{n_2,m}$, related to the $E'$ normal mode are orthogonal among themselves, the relation just stated implies that
\be
\Delta m = m_f-m_i \in \left\{
\begin{array}{lll} 
0, \pm 2, \cdots, \pm \lambda,  && \mbox{ if } \lambda \mbox{ is even} \\
\pm 1, \pm 3, \cdots, \pm \lambda,  && \mbox{ if } \lambda \mbox{ is odd} 
\end{array} \right. .
\label{m-rule}
\ee
Conditions (\ref{n-rule}) and (\ref{m-rule}) are strictly valid only in the harmonic limit and they are easily violated in practice, because the true potential differs from the pure harmonic oscillator. 

The rotational part of the transition matrix elements have to obey the familiar triangular
inequality, 
\be 
\mid J_i-\lambda \mid \le J_f \le J_i+\lambda \qquad \mbox{and} \qquad \pi_f=(-1)^\lambda \pi_i
\ee
together with the summation rule governing the third components both in the laboratory and in the intrinsic frames.

The presence of the $\mathcal{D}_{3h}$ point group symmetry at the endpoint of the group chain in (2) requires extra
care in the definition of the hamiltonian eigenstates and provides also the main source of new selection
rules for electric and magnetic transitions between rotovibrational states by means of the so called {\it vanishing integral rule} \cite{Hou08,Ham,Car97}. The invariance of the rotovibrational hamiltonian under this point group is, in fact, reflected on the requirement that, in the decomposition of the reducible representation of $\mathcal{D}_{3h}$, under which the transition matrix elements transform, the fully symmetric irrep $A_1'$ appears at least once. Said another way:
\be
\Gamma(\Phi_f^E)\otimes\Gamma[\omega_{\lambda\mu}(E,M)]\otimes\Gamma(\Phi_{i}^E) \supset A_1'
\label{times}
\ee
where we have omitted the characters of $\Phi_{n_1}^A$ because they amount to some power of the symmetric representation. If we restrict the above matrix elements to the vibrational ones in the intrinsic frame we get:
\begin{eqnarray}
\langle n_{1f} n_{2f} m_{f} \mid \omega_{\lambda\mu} (R) \mid n_{1i} n_{2i} m_{i}\rangle \ne 0  \\
\leftrightarrow \Phi_{n_{2f}}^{E} \otimes \Gamma[\omega_{\lambda\mu}(E,M)] \Phi_{n_{2i}}^{E} \supset A_1'
\end{eqnarray}

Given that both the rotovibrational states and the electric and magnetic multipole operators transform according to the even bosonic irrep, $A_1'$ or to the odd one, $A_1''$, the vanishing integral rule (\ref{times}), is eventually expressed into the well-known parity rule
\be
 \pi_f=(-1)^{\lambda} \pi_i \;.
\ee
for electric multipole transitions and
\be
 \pi_f=(-1)^{(\lambda-1)} \pi_i \;.
\ee
for the magnetic multipole ones.

\begin{figure}
\begin{center}
\includegraphics[clip=,width=1\textwidth, bb=110 320 500 750]{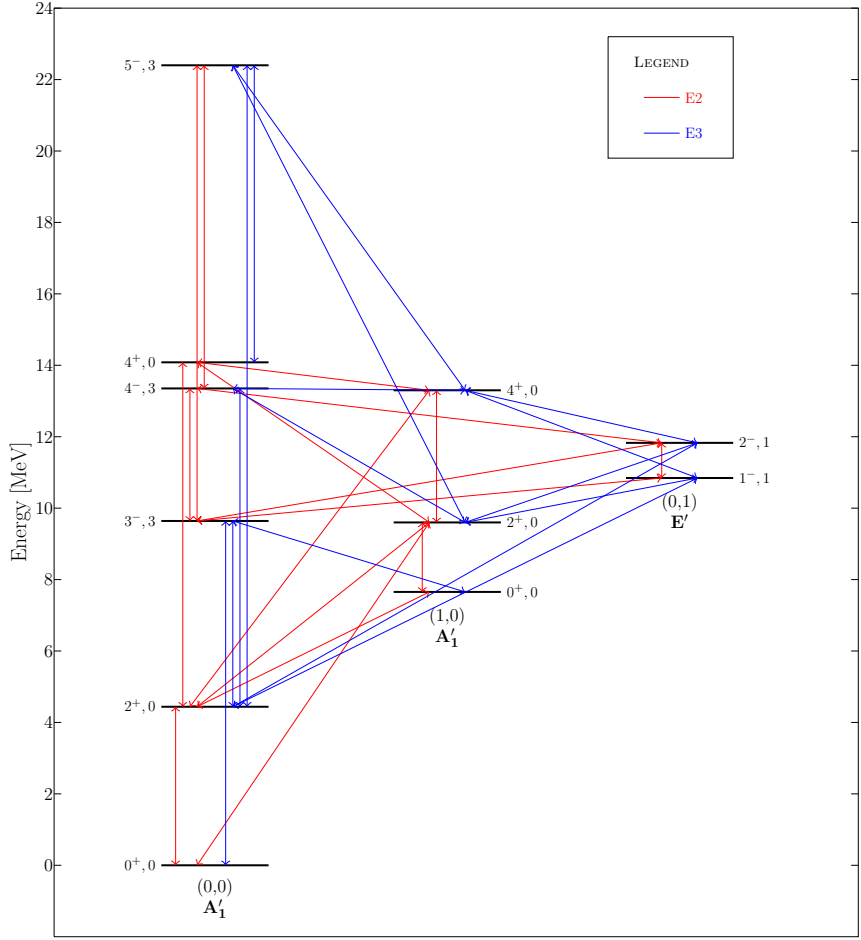}
\end{center}
\caption{Patterns of electric multipole transitions, E2 and E3, in the $\alpha-$cluster model of $^{12}C$.}
\label{Fig:connections}
\end{figure}

Eq. (\ref{times}) governs the vibrational selection rules between bands and together with Table \ref{Tab:oper} allows a first important conclusion: dipole transitions are forbidden. In fact, there is no way to connect any two states with $A_2''$, it would seem like it is possible to connect A-type bands to E-type bands through $E'$, but this is not true and it goes back to the nature of the dipole operator: in fact it can be rewritten in terms of the coordinates of the center of mass and of the equilibrium points without dependence on the normal coordinates. This means that it cannot excite transitions
between intrinsic states of the system.
Thus all the intrinsic states of the system, such as the $3^-$ and $2^+$ states of the g.s. band or the fundamental $0^+$ state and the $1^-$ bandhead of the $E'$ excited band cannot be connected by a dipole transition.
Quadrupole and octupole are instead allowed, it is allowed to go from, let's say, a state with full $A_1'$ symmetry to any other state with the same character because the quadrupole contains $A_1'$ itself (cfr. Fig. (\ref{Fig:connections})). 

\begin{figure}
\begin{center}
\includegraphics[clip=,width=1\textwidth, bb=110 320 500 750]{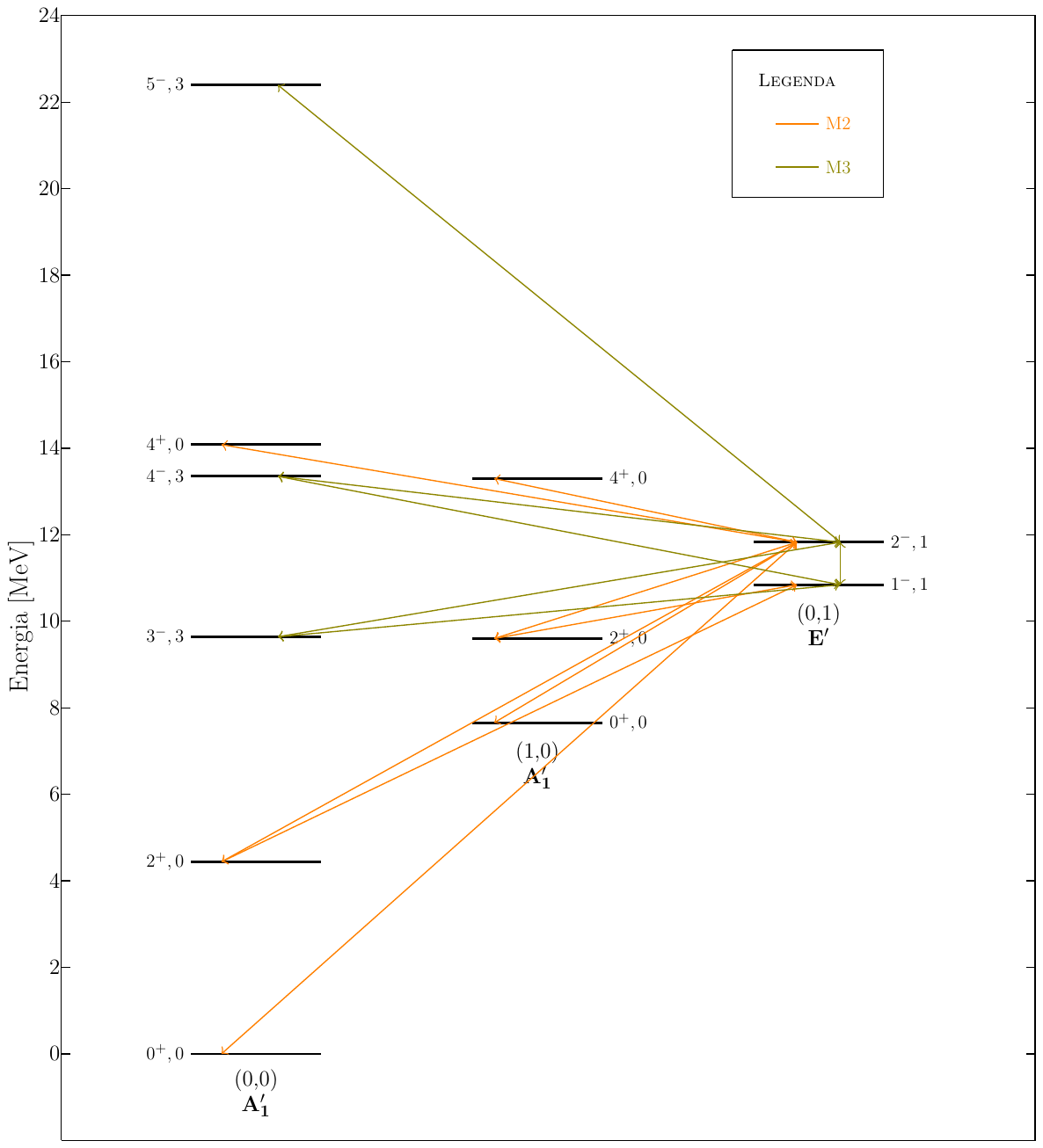}
\end{center}
\caption{Patterns of magnetic multipole transitions, M2 and M3, between observed states that are interpreted within the $\alpha-$cluster model of $^{12}C$. Notice the absence of M3 transitions between bands with the same $A_1'$ symmetry, even when angular momentum couplings would allow them.}
\label{Fig:connectionsM}
\end{figure}

\section{Conclusions}\label{sec5}
The first conclusion is that, whenever discrete point group symmetries are imposed on a certain nuclear system, we must acknowledge that the well-known selection rules that we familiarly use from the spherical shell model or from the collective model have to be supplemented with the requirements coming from the molecular structure and from the $\mathcal{D}_{3h}$ symmetry in this particular case. This bring us to the labeling of states of Table \ref{Tab:states} and to the character of electric and magnetic multipole operators given in Table \ref{Tab:oper} and \ref{Tab:Moper}. Altogether these yield the selection rules discussed in the paper that we summarize in Tab. \ref{Tab:rules}.
\begin{table}[ht]
\begin{center}
\begin{tabular}{ccc}
\hline \hline
$\Gamma(in.) \leftrightarrow \Gamma(fin.)$ & Electric & Magnetic  \\
\hline
$A_1'\leftrightarrow A_1'$ &  E2,3,$\cdots$ & M4,6,7,$\cdots $ \\
$E' \leftrightarrow  E'$ &   E2,3,$\cdots$ & M2,3,$\cdots $ \\
$A_1'\leftrightarrow E'$ &   E2,3,$\cdots$ & M2,3,$\cdots$ \\
\hline \hline
\end{tabular}
\end{center}
\caption{Summary of allowed electric and magnetic multipole transitions between bands of a system with $\mathcal{D}_{3h}$ symmetry.}
\label{Tab:rules}
\end{table}

To this table one should add a selection rule on the vibrational quantum numbers discussed above, only in the case of strict adherence to the harmonic approximation, but if the true potential cannot be taken as harmonic, then transitions are possible also between bands with larger differences in the vibrational quantum numbers. While the last rule can be taken as an indication and it might be easily violated in practice, the rules in Table \ref{Tab:rules} are inherent in the model.  

One should notice that, although bands are properly named after their vibrational character, the full roto-vibrational states have either $A_1'$ or $A_1''$ character depending on their parity. One can see that dipole transitions (E1) between $\alpha-$cluster states are forbidden and that the lowest allowed electric multipole is the quadrupole (E2). We have summarized in Fig. (\ref{Fig:connections}) the lowest allowed electric transitions and in Fig. (\ref{Fig:connectionsM}) the lowest allowed magnetic transitions.
Analogously the magnetic dipole (M1) is forbidden. It is interesting to note that the two lowest experimentally measured $1^+_1$ and $1^+_2$ states in $^{12}$C at about 12.71 and 15.11 MeV are indicated as non-cluster states in the analysis of Ref.\cite{MLB14}, where it is said that no $1^+$ state can be formed in the 3$\alpha$-cluster model. This is not completely true. In fact a single $1^+$ state appears in Table \ref{Tab:states} in the $n_2=3$ band \footnote{There is another in the $n_2=5$ band, cfr. Ref.\cite{Stellin15}}. However this state should be found at some higher energy (and possibly in some energy regime where the importance of clusters has already disappeared) inside a band with $0^+, 1^\pm, 2^{\pm,+}, \cdot$, but the real reason why these are most certainly not cluster states is that they do decay by M1 transitions and these are forbidden in the model. They necessarily have some different nature. 

Another conclusion is that M2, M3 and M5 transitions are forbidden between bands with $A_1'$ character, while they are allowed in other cases. If these weak transitions could be measured (and unfortunately this is often not the case) they could serve as robust signatures.


\section{Acknowledgments}
We acknowledge private communication with A. Gelessus (Jacobi Univ., Bremen) about selection rules for electric multipoles in molecules.

\section{References}
\bibliography{Bibliography}{}
\bibliographystyle{plain}


\end{document}